\documentclass[manuscript]{aastex}
\usepackage{placeins}

\slugcomment{}

\shorttitle{Light-Echo Distance and Structure for T Pyx}
\shortauthors{Sokoloski et al.}

\begin{document}



\title{The Recurrent Nova T Pyx: Distance and Remnant Geometry from
  Light Echoes}


\author{J. L. Sokoloski\altaffilmark{1}, Arlin P. S. Crotts, and
  Helena Uthas\altaffilmark{2}} 

\affil{Columbia Astrophysics Laboratory, Columbia University, 550 West
  120th Street, New York, NY 10027, USA}

\and

\author{Stephen Lawrence\altaffilmark{3}}
\affil{Department of Physics and Astronomy, Hofstra University,
  Hempstead, NY 11549, USA}


\begin{abstract} 

The recurrent nova T~Pyxidis (T~Pyx) is well known for its small
binary separation, its unusually high luminosity in quiescence, and
the spectacular Hubble Space Telescope ($HST$) images of its
surrounding remnant.  In 2011 April, T~Pyx erupted for the first time
since 1966.  Here we describe $HST$ observations in late 2011 of a
transient reflection nebula around the erupting white dwarf (WD).  Our
observations of this light echo in the pre-existing remnant show that
it is dominated by a clumpy ring with a radius of about 5\arcsec\, and
an inclination of 30$^\circ$ -- 40$^\circ$, with the eastern edge
tilted toward the observer.  The delay times between the direct
optical light from the central source, and the scattering of this
light from dust in several clumps with the same foreground distance as
the central source, give a distance to T~Pyx of $4.8\pm0.5$~kpc.
Given past evidence from two-dimensional optical spectra that the
remnant contains a shell-like component, it must actually consist of a
ring embedded within a quasi-spherical shell.  The large distance of
4.8~kpc supports the contention that T~Pyx has an extraordinarily high
rate of mass transfer in quiescence, and thus that nova explosions
themselves can enhance mass loss from a donor star, and reduce the
time between eruptions in a close binary.

\end{abstract}





\section{Introduction}

During a nova eruption, material is explosively ejected from the
surface of an accreting white dwarf (WD). Nova explosions thus alter
the evolution of the WD.  Copious emission from novae also probably
influences the companion star and the host cataclysmic variable
\citep[CV; e.g.,][]{patterson2012}.  The companion star, in turn, must
modify the shape of the ejecta.  However, the way in which the
exploding WD and the donor star disturb one another --- e.g., the
degree to which the close companion generates asymmetry in the
remnant, the mechanism of such shaping, and the affect of the outburst
on the mass transfer rate --- are poorly understood.

T~Pyx is an unusual CV that from years 1890 and 1966 suffered
recurrent nova eruptions roughly every 20 years. Compared to other CVs
with similar orbital periods, T~Pyx has a much higher accretion rate,
and hence luminosity \citep[e.g.,][]{patterson1998}.  It can also have
a high ejecta mass for a recurrent nova
\citep{nelson2013,patterson2013}.  Both the high luminosity and
frequent nova eruptions may be part of a self-perpetuating cycle in
which nova eruptions trigger enhanced mass-loss from the donor star,
as first proposed by \citet{knigge2000}.  Previous images of the
remnant around T Pyx in H$\alpha$ and~[NII] line emission revealed
thousands of knots within about 6\arcsec\,~of the central binary and a
halo possibly extending out to 10\arcsec\,~--- resulting from multiple
recent nova eruptions \citep{shara1989,shara1997,schaefer2010}.  The
most recent eruption was discovered on 2011 April 14.29
\citep{waagen2011}.  Because T~Pyx is the only nova confirmed to have
erupted within a pre-existing remnant
\cite[e.g.,][]{shara1989,shara1997,schaefer2010}, it provides a rare
opportunity to examine the ejecta from a nova using a light
echo.\footnote{N Sagittarii 1936 may have also produced such an echo
  \citep{swope1940}.  The light and ionization echoes from GK~Persei,
  V458~Vulpeculae, and V838~Monocerotis probed either surrounding
  planetary nebulae or pre-existing circumstellar dust from an unusual
  type of stellar eruption
  \citep{bode2004,wesson2008,bond2003,soker2007}.}

Below, we describe how Hubble Space Telescope ($HST$) observations of
the light echo around T Pyx reveal: 1) how much the companion has
shaped the remnant; and 2) T~Pyx's distance, with implications for
understanding how novae influence binary stellar evolution.

\section{Observations and Data Reduction}

%

$HST$ observed T Pyx using the Wide Field Camera 3 (WFC3) nine times
between 2011 July and 2012 May, and using the Space Telescope Imaging
Spectrograph (STIS) with G430L and G750L low-resolution gratings and a
slit width of 2\arcsec\, three times between 2011 May and 2011
December (program 12448; see Table~1). To isolate the signal due
primarily to scattering by dust, we selected four filters that exclude
strong recombination lines: F225W (UV wide filter), FQ422M (blue
continuum), F547M (Str{\"o}mgren-$y$), and F600LP (long-pass).  Every
WFC3 observation included F547M exposures, which we refer to as {\em
  visual-light} observations.

To reduce the WFC3 images, we subtracted the point spread function
(PSF) of the central source from 2011 July, and interpolated across
small regions where charge was saturated due to PSF brightness and
column bleeding.  For F547M images, we define an angular distance from
the central binary, $\theta_{\rm bright}$, as that within which
residuals from the subtracted central-source PSF reduced sensitivity
to the light echo.  In 2011 July, $\theta_{\rm bright}$ was greater
than the size of most of the known remnant (approximately 6\arcsec).
It decreased in later epochs as the central source faded.

\begin{deluxetable}{cllllcllr}
\tabletypesize{\footnotesize}
\tablecaption{Log of HST observations \label{tab:obslog}} 
\tablewidth{0pt}
\tablehead{
\colhead{Epoch} & \colhead{Instru-} &
\colhead{Date} &
\colhead{Day\tablenotemark{a}} & \colhead{Exp} & \colhead{Apertures} &
\colhead{Filters/} & \colhead{Offset\tablenotemark{c}} &
\colhead{PA\tablenotemark{d}} \\ 
\colhead{} & \colhead{ment} & 
\colhead{} &
\colhead{} & \colhead{Time\tablenotemark{b}} & \colhead{} &
\colhead{Gratings} & \colhead{(\arcsec)} & \colhead{($^\circ$)}
}
\startdata
1 & STIS & 2011-05-19 & 35.1 & 205.0 & STIS-52X2 & G750L & -1.229 & 61.05 \\        
 &  &  & 35.1 & 615.0 & STIS-52X2 & G750L & -1.229 & 61.05 \\        
 &  &  & 35.1 & 205.0 & STIS-52X2 & G430L & -1.229 & 61.05 \\        
 &  &  & 35.1 & 615.0 & STIS-52X2 & G430L & -1.229 & 61.05 \\        
 &  &  & 35.2 & 205.0 & STIS-52X2 & G750L & 1.303 & 57.55 \\         
 &  &  & 35.2 & 615.0 & STIS-52X2 & G750L & 1.303 & 57.55 \\        
 &  &  & 35.2 & 205.0 & STIS-52X2 & G430L & 1.303 & 57.55 \\        
 &  &  & 35.2 & 615.0 & STIS-52X2 & G430L & 1.303 & 57.55 \\        
 &  &  & 35.2 & 820.0 & STIS-52X2 & G750L & 3.654 & 52.65 \\        
 &  &  & 35.2 & 820.0 & STIS-52X2 & G430L & 3.654 & 52.65 \\        
2 & STIS & 2011-05-26 & 42.1 & 820.0 & STIS-52X2 & G750L & -4.341 & 67.85 \\        
 &  &  & 42.1 & 820.0 & STIS-52X2 & G430L & -4.341 & 67.85 \\       

3 & WFC3  & 2011-07-08 & 85.7 & 300.0 &  UVIS-QUAD-SUB &
FQ422M & & \\
 &  &  &   & 500.0 &  UVIS2-2k2C-SUB &
F547M & &\\
 &  &  &   & 1197.0 & UVIS2-2k2C-SUB &
F225W & & \\
4 & WFC3 & 2011-09-19 & 158.7 & 300.0 & UVIS-QUAD-SUB &
FQ422M & & \\
 &  &  &  & 500.0 & UVIS2-2k2C-SUB & F547M & & \\ 
 &  &  &  & 800.0 & UVIS2-2k2C-SUB & F225W & & \\
 &  &  &  & 40.0 & UVIS2-2k2C-SUB & F600LP & & \\
5 & WFC3 & 2011-09-26 & 165.1 & 1125.0 & UVIS-QUAD-SUB & FQ422M & & \\
 &  &  &  & 900.0 & UVIS2-2k2C-SUB & F547M & & \\
6 & WFC3 & 2011-11-16 & 215.8 & 1125.0 & UVIS-QUAD-SUB & FQ422M & & \\
 &  &  &  & 900.0 & UVIS2-2k2C-SUB & F547M & & \\
\tablebreak
7 & WFC3 & 2011-11-25 & 224.8 & 300.0 & UVIS-QUAD-SUB & FQ422M & & \\
 &  &  &  & 500.0 & UVIS2-2k2C-SUB & F547M & & \\
 &  &  &  & 800.0 & UVIS2-2k2C-SUB & F225W & & \\
 &  &  &  & 40.0 & UVIS2-2k2C-SUB & F600LP & & \\

8 & STIS & 2011-12-05 & 235.6 & 830.0 & STIS-52X2 & G750L & 2.000 & -119.35 \\      
 &  &  & 235.6 & 830.0 & STIS-52X2 & G430L & 2.000 & -119.35 \\      
 &  &  & 235.7 & 830.0 & STIS-52X2 & G750L & -2.000 & -91.35 \\      
 &  &  & 235.7 & 830.0 & STIS-52X2 & G430L & -2.000 & -91.35 \\
 &  &  & 235.5 & 830.0 & STIS-52X2 & G750L & -5.300 & -109.35 \\      
 &  &  & 235.6 & 830.0 & STIS-52X2 & G430L & -5.300 & -109.35 \\      
 &  &  & 235.5 & 830.0 & STIS-52X2 & G750L & 4.500 & -109.35 \\      
 &  &  & 235.5 & 830.0 & STIS-52X2 & G430L & 4.500 & -109.35 \\      

9 & WFC3 & 2011-12-10 & 240.7 & 2490.0 & UVIS2-2k2C-SUB & F547M & & \\
10 & WFC3 & 2012-01-22 & 282.8 & 5295.0  & UVIS2-2k2C-SUB & F547M & & \\  
11 & WFC3 & 2012-02-22 & 314.2 & 5295.0 & UVIS2-2k2C-SUB
& F547M & & \\ 
12 & WFC3 & 2012-05-21 & 403.4 &  5295.0 & UVIS2-2k2C-SUB 
& F547M & & \\ 
\enddata
\tablenotetext{a}{Days after $t_0 =\, $JD~2455665.79.}
\tablenotetext{b}{Exposure time, in seconds.}
\tablenotetext{c}{Offset of the central source perpendicular to the
  center of the STIS
  slit, in the red-ward dispersion direction.}
\tablenotetext{d}{The position angle of the 
slit, projected onto
  the sky, east of north.}
\end{deluxetable}
\FloatBarrier

To reduce the STIS spectra, we followed the standard CALSTIS pipeline
within IRAF.  We also manually removed the extensive ``warm pixel''
artifacts, cosmic rays and their charge-transfer tails through careful
inspection of the three CR-SPLIT sub-exposures taken in each spectrum.

\section{Results: Variable Ring of Emission \label{sec:results}}

\begin{figure}
\begin{minipage}{6.5in}
\vspace*{-20cm}
\includegraphics[width=6.5in]{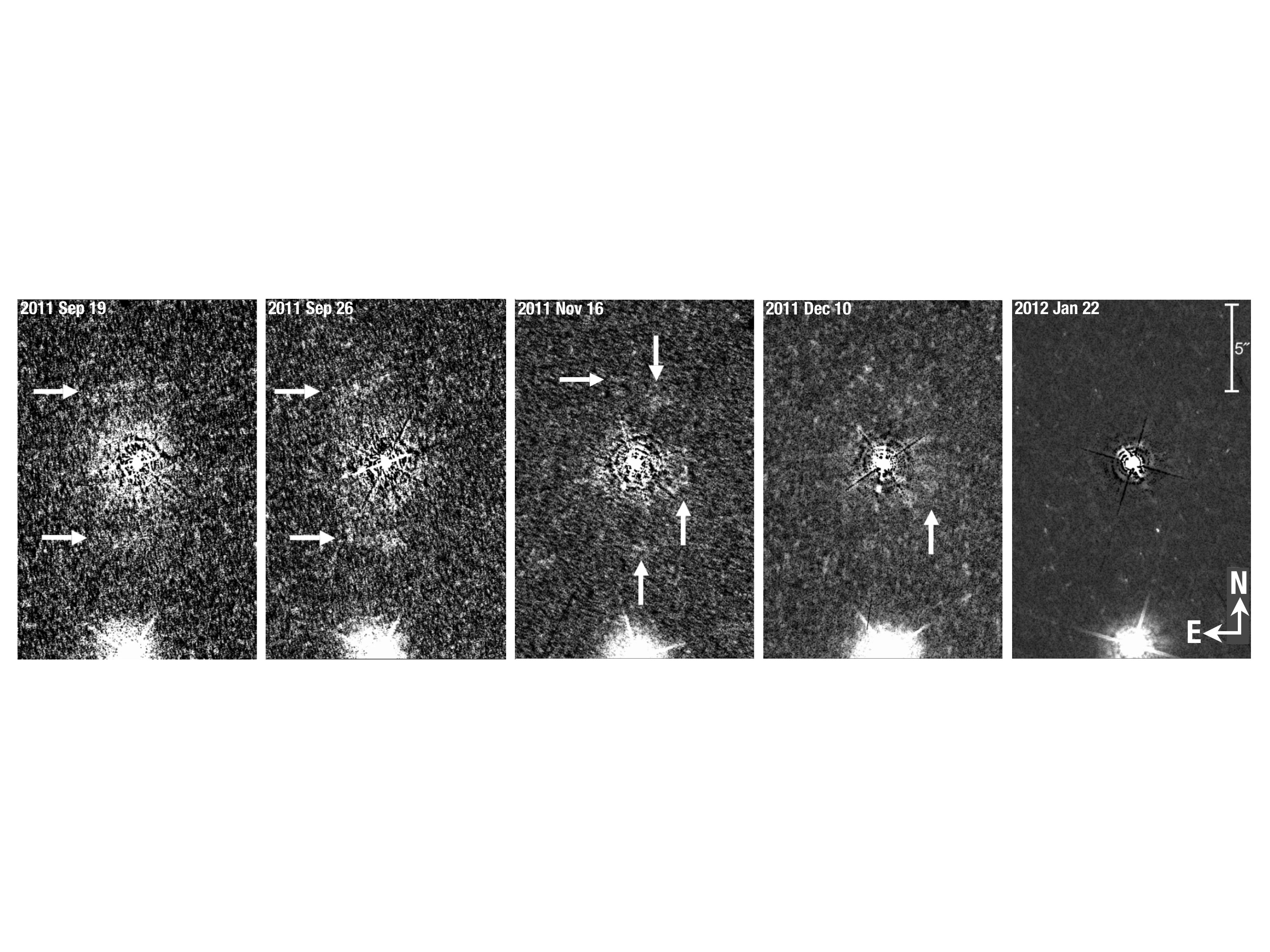}
\vspace{-4.5cm}
\caption{Series of $HST$/WFC3 images, from 2011 Sep 19, Sep 26,
  November 16, December 10, and 2012 January 22, showing the
  visual-light echo.  In the left two panels, the top and bottom
  arrows point to echos that appeared in regions N1 and S1 (as defined
  in Figure~\ref{fig:echoLCs}), respectively. The arrows in the center
  panel indicate echos that appeared in regions N2, NW, W, and S2
  (clockwise from top).
\label{fig:ims}}
\end{minipage}
\end{figure}

Between 2011 September and December (5 to 8 months after the
eruption's start), $HST$/WFC3 detected multiple transient patches of
visual-light emission in a ring-like configuration around the central
binary (see white arrows in Figure~\ref{fig:ims}). We refer to this
transient emission as an {\em echo} and justify this choice of
terminology in \S\ref{sec:strucanddist}.  In 2011 July, the central
source was too bright for echoes to be detectable at the location of
the known remnant. In 2011 September, the echo was dominated by two
swathes of emission 4\arcsec\, to 6\arcsec\, to the north and south of
the central binary, respectively (see Figure~\ref{fig:ims}; $4\arcsec
< \theta < 6\arcsec\,,$ where $\theta$ is the angular distance from
the central binary).  By 2011 November, the echo had shifted
dramatically --- the brightest portion lay to the west of the central
binary, with angular extent $\theta \lesssim 4\arcsec$, and an
additional new patch of emission was present to the northwest of the
central source.  Several small patches of emission also remained to
the north and south of the central binary, but at larger $\theta$ than
in 2011 September.  In 2011 December, the echo continued progressing
west and to slightly larger $\theta$, while fading overall.  Virtually
all of the visual echo had disappeared by 2012 January 22, within 10
months of the eruption's start.  $HST$/WFC3 detected no significant
visual echo in 2012 March or May.  $HST$/WFC3 also detected no
extended structure through filters F225W or FQ422M during any epoch of
this observing campaign, and only hints of a signal at the location of
the visual echo with the F600LP filter.\footnote{The lack of a F600LP
  exposure in 2011 July, and the corresponding lack of a high-quality
  empirical PSF at these wavelengths, prevented us from obtaining
  quantitative results for the red echo.}  Figure~\ref{fig:ims} shows
the visual echo traveling predominantly from east to west around an
irregular, patchy partial ring.

In addition to the partial ring, $HST$/WFC3 detected several other
clumps of reflecting material.  Two small patches of visual emission
appeared briefly to the southeast of the ring and central binary ---
one in 2011 September at $\theta \approx 12\arcsec$ and $PA \approx
130^\circ$, and one in 2011 November at $\theta \approx 10\arcsec$ and
$PA \approx 137^\circ$.  Given the low signal-to-noise level of the
visual echoes, we cannot, however, place strong constraints on
additional material outside roughly $\theta \approx 6\arcsec$.  Inside
the ring, a diffuse patch was present at $PA \approx 35^\circ$ in 2011
September.  Therefore, between 2011 September and December, $HST$
detected visual-light echoes from regions associated with most of the
dominant features of the quiescent-state, H$\alpha$-emitting remnant
\citep[as shown in, e.g.,][]{shara1989,schaefer2010}.

\begin{figure}
\begin{center}
\includegraphics[width=12cm]{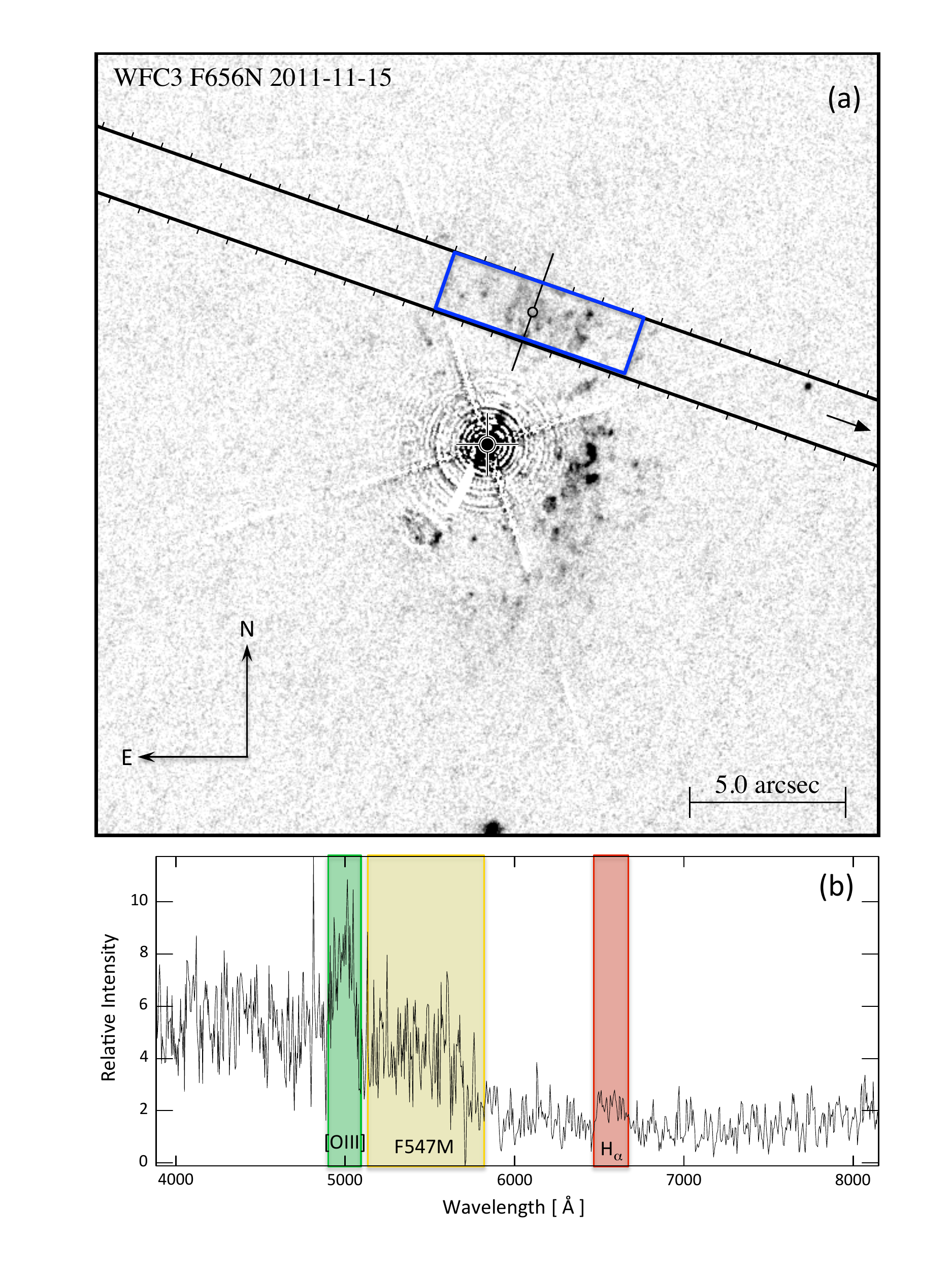}
\end{center}
\vspace{-1.2cm}
\caption{\footnotesize STIS spectrum of the echoing remnant. {\em a)}
  The position of the slit on 2011 December 5 superimposed on a
  PSF-subtracted archival WFC3/F656N image from 2011 November 15
  \citep[for a description of the WFC3/F656N observations,
    see][]{shara2013}.  The blue rectangle shows the region of the
  spectral extraction.  {\em b)} A summed, one-dimensional extraction
  of the combined G430L and G750L spectra.  The left-most rectangle
  shows the spectral range of diffuse, low-velocity [O III]~5007
  emission uniformly filling the slit; the right-most rectangle shows
  the same for H$\alpha$ emission.  STIS detected line emission
  clearly in [O~III] and weakly in H-alpha.  The central rectangle
  shows the FWHM bandpass of the WFC3/F547M filter used in our
  light-echo analysis; there is no strong line emission in this
  passband.  The echo that we detected is thus consistent with
  dust-scattered continuum from the outburst.
\label{fig:spec}}
\end{figure}

The remnant's spectrum was dominated by two emission lines: H$\alpha$
and [O\,III].  Although our spectroscopic observations in 2011 May,
and two slit positions in 2011 December, furnished no useful
information due to high background from the PSF of the bright central
source, and a third slit position in 2011 December covered no obvious
echo patches, the 2011 December slit position with an offset of
4.5\arcsec\, included the northwest echo region, from which we
detected WFC3/visual-band emission on both 2011 November 15 and 2011
December 10.  The long-wavelength (G750L) spectrum at this position
revealed clear H$\alpha$ line emission with a complex profile most
likely due to the distribution of material within the slit
(Figure~\ref{fig:spec}).  In the short-wavelength spectrum (G430L),
the most obvious signal was emission from [O~III]~$\lambda$5007\AA.
No strong emission lines appeared in the wavelength region to which
the F547M filter is sensitive --- consistent with the visual echo
consisting of predominantly scattered continuum emission.

\FloatBarrier

\section{Structure of the Remnant and Distance to
  T~Pyx\label{sec:strucanddist}}  

%
%

The most natural interpretation of the variable visual emission from
the old remnant around T~Pyx is that light emitted by the WD near the
peak of the 2011 nova eruption was reflected into our line of sight by
dust in the remnant.  Numerous pieces of evidence support this
interpretation.  Using past estimates of T~Pyx's distance of
$3.5\pm1$~kpc \citep{patterson1998,schaefer2010,shore2011}, the delay
between the peak of the direct optical emission and the appearance of
the echo was consistent with the light-travel time from the central
binary to the remnant.  Furthermore, dust exists in the remnant
\citep[with a dust-to-gas ratio of approximately $2 \times 10^{-4}$ by
  mass;][]{contini1997}.  Additionally, although the remnant was flash
ionized \cite[][]{shara2013}, radiation from recombination did not
contribute significantly to the visual echo, as evidenced by: 1) the
low level of the Paschen recombination continuum based on a scaling
from the H$\alpha$ flux; and 2) our non-detection of any strong
recombination lines (e.g., [O\,I] 5577\AA\, or He\,I~5876\AA) in the
spectrum in Figure~\ref{fig:spec}.  Our non-detection of echoes in the
F225W and FQ422M filters and detection of echoes in the F547M and
F600LP filters is consistent with the colors of emission from the
central source near optical maximum \citep[see][and assuming gray
  dust]{surina2013}.  Finally, the amplitude of fluctuations in
brightness of the extended optical light was comparable to that from
the central source.  We thus refer to the transient, extended
visual-band emission as an {\em echo}.

The delay times between variations in brightness of the central source
and those of the various echo patches reveal the geometry of the
remnant around T~Pyx.  Light from the central source reflecting off
dust in the remnant travels to us via longer paths than light coming
directly from the central source; the difference in arrival times
between two photons emitted simultaneously but traveling these
different paths --- the {\em delay time}, $t$ --- depends upon the
reflecting parcel's location with respect to the illuminating central
star.  Following \cite{couderc1939}, the distance from the WD to the
reflecting parcel projected along our line of sight (the {\em
  foreground distance}, $z$) is given by
\begin{equation} \label{eqn:echoeqn}
z = \frac{\rho^2}{2ct} - \frac{ct}{2},
\end{equation} 
where $\rho = \theta D$ is the distance between the central source and
reflecting parcel projected onto the plane of the sky (with $D$ the
distance from the observer to T~Pyx), $c$ is the speed of light, with
$z$ measured from the plane of the sky containing the central source
{\em toward} the observer.  Table~\ref{tab:delaytimes} lists the delay
times that minimize $\chi^2$ when we model the light curves from the
six regions that have the strongest echo signals as shifted versions
of the optical light curve from the erupting WD \citep[using $V$-band
  observations from the American Association of Variable Star
  Observers; AAVSO;][]{henden2013}.  Since $\theta$ is observable, for
a given $D$ the delay times provide the 3-dimensional geometry of the
reflecting material.  Conversely, if $z$ is known for even a single
reflecting parcel, the delay time for that parcel gives $D$.

\begin{deluxetable}{lcll}
\tablecaption{Delay Times and Foreground Distances for Key
    Echo Regions\label{tab:delaytimes}}   
\tablewidth{0pt}
\tablehead{
\colhead{Location of} & \colhead{Delay Time\tablenotemark{b}} &
\colhead{$\theta$\tablenotemark{c}} & \colhead{$z$\tablenotemark{d}}\\
\colhead{Echo Patch\tablenotemark{a}} & \colhead{(day)} & \colhead{(\arcsec)} &
\colhead{($10^{17}$ cm)}
}
\startdata
north (N1) & $125\pm8$ & 4.52 & $0.0\pm0.4$ \\ 
south (S1) & $136\pm5$ & 4.99 & $0.1\pm0.4$\\ 
north \#2 (N2) & $156\pm5$ & 5.50 & $-0.1\pm0.4$\\ 
south \#2 (S2) & $162\pm3$ & 5.82 & $0.0\pm0.4$\\ 
northwest (NW) & $159^{+10}_{-1}$ & 3.99 & $-1.1\pm0.2$ \\ 
west (W) & $164^{+20}_{-5}$ & 3.32 & $-1.5\pm0.2$\\ 
\enddata
\tablenotetext{a}{Figure~\ref{fig:echoLCs} shows the locations of the
  echo patches.}
\tablenotetext{b}{
Uncertainties correspond to the range of delay times immediately
around the optimal $t$ providing acceptable matches between the light
curves of the central source and the echo patch.  }
\tablenotetext{c}{$\theta$ is the angular distance between the
  central binary and the center of light of the echo 
  patch.}
\tablenotetext{d}{Foreground distances, $z$, are based upon a
  distance to T~Pyx of $4.8\pm0.5$~kpc.
} 
\end{deluxetable}

\begin{figure*}
\hspace{-0.3cm}
\includegraphics[width=9.6cm]{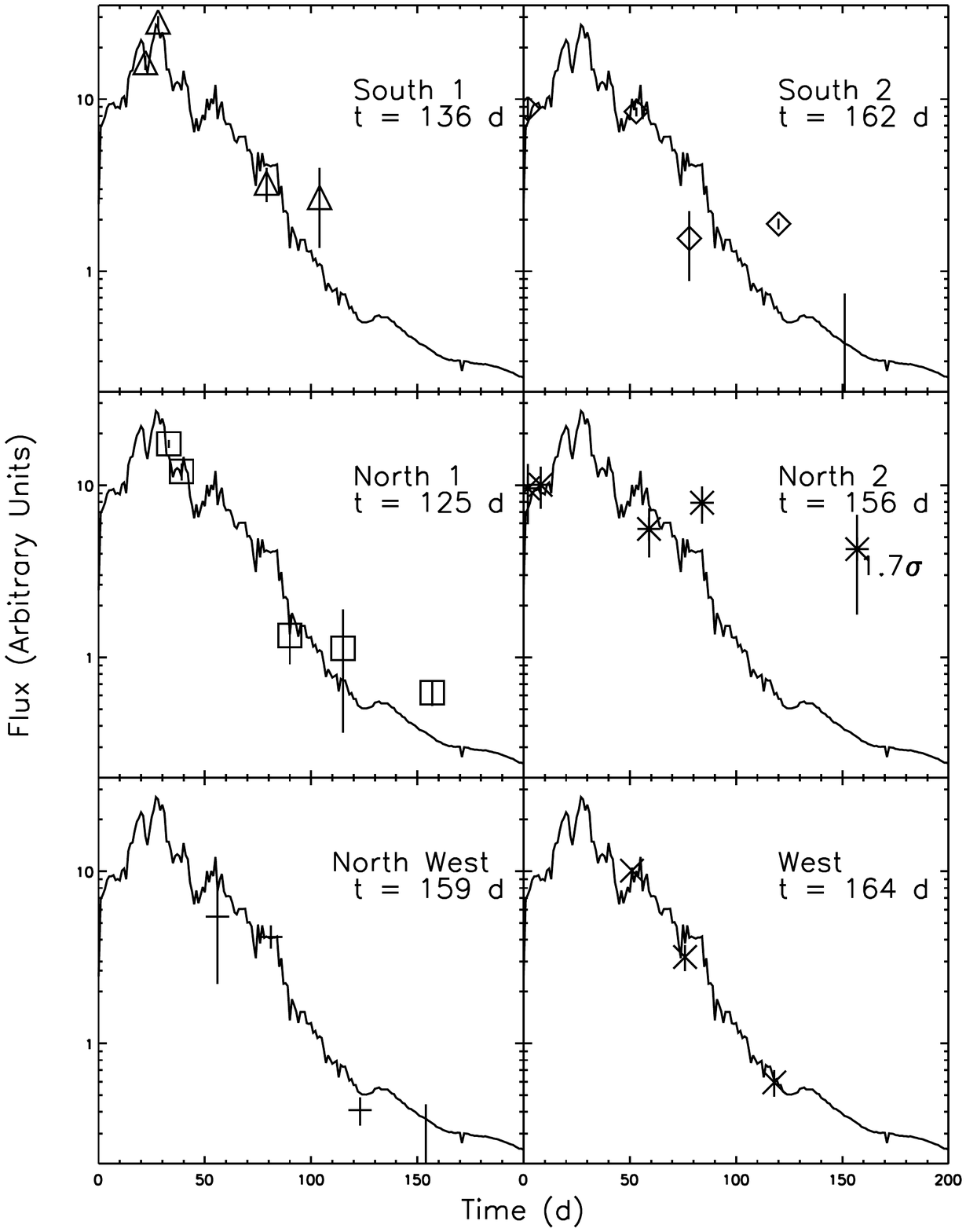}
\hspace{-0.3cm}
\includegraphics[width=6.9cm]{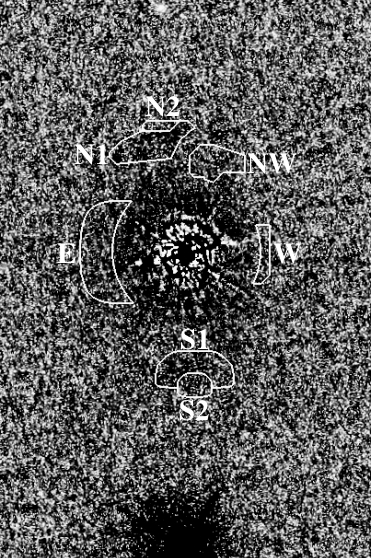}
\caption{{\em Left panel:} $V$-band light curves of the erupting WD,
  from the AAVSO \citep{henden2013}, with visual-band
  light curves from the six strongest echo regions shifted by 
  the delay times listed in Table~\ref{tab:delaytimes} and 
over-plotted.  {\em Right panel}: Extraction regions for the echo
light curves, superimposed on the PSF-subtracted WFC3 visual image
from 2011 July. 
\label{fig:echoLCs}}
\end{figure*}

Even before we consider the precise delay times, the morphology and
progression of the light echo reveals that the principal structure of
the remnant around T~Pyx is a clumpy ring that is modestly inclined
with respect to the plane of the sky.  Recognizing that the brightness
of the central source did not allow for detection of echoes with delay
times of less than about 120 days, the gross motion of the echo across
the remnant in a westerly direction demonstrates that foreground
distance decreases to the west.  The lack of a detectable echo peak
from the eastern side of the remnant (region E in
Figure~\ref{fig:echoLCs}) after 2011 September is consistent with
material on that side producing only reflections with short delay
times and therefore having larger foreground distances than material
on the western side of the remnant.  In addition, although the echo
progressed to larger $\theta$ (especially in the north and south
between 2011 September and December), it did not subsequently move
inward, as would be expected if the main reflecting structure had a
quasi-spherical shell-like shape.  Thus, the more-or-less circular
appearance of the known remnant is due not to spherical symmetry, but
to a 2D disk-like structure. Moreover, the plane containing this
structure is rotated with respect to the plane of the sky around a N-S
axis.  The consistency between the locations of the echoes and the
known remnant indicates that the remnant is the source of the
reflecting material, and that this material sits in a ring.

Assuming the binary is coplanar with the ring, the delay times for the
echo regions to the north and south of the central binary give the
distance to T~Pyx.  The assumption that the binary is coplanar with
the inclined ring is justified on multiple grounds.  If the ring was
in front of the binary from the point of view of the observer,
symmetry arguments suggest that we would expect another ring behind
the binary generating an echo at late times.  But no such second echo
occurred, which rules out values of $D$ larger than $4.8$~kpc, derived
below.  Distances less than 4.8~kpc are disfavored by
\citet{shore2011}.  Moreover, the ring's inclination around the N-S
axis means that all reflecting clumps to the north and south must have
the same $z$, which only happens for a distance of around 4.8~kpc.
With that distance, the central binary does indeed lie near the center
of the ring (i.e., $z \approx 0$ for regions N1, N2, S1, and S2).
Plugging $\theta$ and $t$ for the four echo patches to the north and
south of the central binary into Equation~\ref{eqn:echoeqn}, we obtain
four distinct distance estimates of 4.8, 4.7, 4.9, and 4.8~kpc.
Although the dispersion in the distance measurements from the four
light-echo patches is nominally just 0.1~kpc, additional uncertainty
of $\sim10$\% associated with our geometrical assumptions and the
complexity of the central-source light curve indicate $D = 4.8 \pm
0.5$~kpc.

The foreground distances to the western and northwestern clumps (see
Table~\ref{tab:delaytimes}) give inclination angles for the ring of
$30^\circ$ and $40^\circ$, respectively.  For comparison, published
estimates of the inclination of the central binary range from
$6^\circ$ to about $30^\circ$
\citep[e.g.][]{shahbaz1997,patterson1998,selvelli2008,uthas2010,patterson2013}.

H$\alpha$ observations with $HST$/F656N \citep[][]{shara2013} support
our interpretation of the light echo.  If the transient H$\alpha$
emission from the remnant was due to recombination following
ionization by the UV flash at the start of the eruption, and the
visual-band echo was due to dust-scattering of the optical emission
that peaked about 30 days into the eruption (as we propose here), we
would expect the H$\alpha$ brightening to lead the visual echo by
about 30 days.  Thus, we expect the maximum angular extent of the
H$\alpha$ emission to exceed that of the visual echo at the same
observational epoch by $\Delta \theta = (c\, \Delta t / D) =
1.1\arcsec\, (D/4.8~{\rm kpc})^{-1}\,(\Delta t/30~{\rm d})$, where
$\Delta t$ is the difference between the delay times for the H$\alpha$
and visual brightenings, along directions where $z=0$.  Averaging over
the northern and southern echo patches in 2011 September, the
H$\alpha$ emission described by \citet{shara2013} is approximately an
arcsec more extended in $\theta$ than the visual echo (and more to the
west) --- supporting both the physical mechanism of the echo and our
distance estimate.  The east-to-west motion of the H$\alpha$
recombination emission, and lack of a connecting arc between the
northern and southern H$\alpha$ emission components in 2011 September,
are also consistent with the reflecting structure being dominated by a
patchy, inclined ring.

\section{Implications}

The finding of a ring-like component in the remnant around T~Pyx, with
an inclination comparable to that of the orbital plane of the binary,
suggests that the companion helped shape the ejecta.  Account for the
optical spectra of \citet{obrien98}, it appears that the ring-like
formation that produced the visual echo is part of a three-dimensional
shell-like structure.  Moreover, using the distance of
$4.8\pm0.5$~kpc, the radial velocity of about 500~km~s$^{-1}$ that
\citet{obrien98} found for material at the back and the front of the
shell is comparable to the expansion speeds in the plane of the sky of
many of the knots with $\theta$ between about 4\arcsec\, and
5\arcsec\, \citep[the expansion speeds in the plane of the sky of the
  fastest, outermost knots are close to $1,000\;{\rm km}\,{\rm
    s}^{-1}\{D/4.8\,{\rm kpc}\}$;][]{schaefer2010}.  Although the
mechanism by which the donor star might shape the remnant is poorly
understood, hydrodynamical simulations demonstrate that an orbiting
companion can produce rings such as those seen in images of old novae
\citep{lloyd1997,porter1998}.  Alternatively, we cannot rule out the
possibility that at least some of the material in the torus emanated
directly from the heated donor star.  In T~Pyx, the increase in
orbital period during quiescence indicates than even between
outbursts, the binary blows a strong wind \citep[e.g.,][]{uthas2010}.

A distance to T~Pyx of $4.8\pm0.5$~kpc means that the quiescent-state
accretion rate in this system is inescapably high.  The new, more
accurate distance also reduces the uncertainties on all of the
fundamental parameters of this system.  Using this distance, the
accretion rate is confirmed to be approximately three orders of
magnitude higher than expected for mass transfer driven by
gravitational radiation \citep[e.g.,][]{knigge2000}.  If such high
accretion rates can be stimulated by nova eruptions in close binaries
\citep[as suggested by][]{knigge2000}, and if they can be maintained
for even a few decades following an explosion, then novae must
strongly impact the evolution of a close binary.  By supporting the
idea that a nova can catapult a CV into a state of rapid mass
transfer, during which the WD can in principle gain mass efficiently
\citep{bours2013}, our confirmation of the large distance and high
quiescent-state rate of accretion for T~Pyx could have bearing on the
problem of how the WDs in CVs can grow to surprisingly large masses
\citep{zorotovic2011}.  And although the WD in T~Pyx does not appear
to be on its way to the Chandrasekhar limit
\citep{patterson2013,nelson2013}, its illustration of the phenomenon
of nova-stimulated mass transfer could provide a hint about how others
CVs might possibly evolve to become type Ia supernovae.

\acknowledgments


We are grateful to J. Patterson, D. Zurek, C. Knigge, M. Shara, and
B. Schaefer for useful discussions.  We appreciate STScI's
(particularly Tony Roman's) patience with these difficult
observations.  J.L.S., A.C., and H. U. acknowledge support from grant
HST-GO-12448.  J.L.S. acknowledges support from NSF award AST-1211778.



{\it Facilities:} \facility{HST (WFC3, STIS)}, \facility{AAVSO}.

\clearpage



\clearpage

\end{document}